\documentclass[aps,prl,reprint,showpacs,superscriptaddress]{revtex4-1}

\usepackage{graphicx}
\usepackage{amsmath}
\usepackage{amssymb}
\usepackage{epsfig}

\renewcommand{\v}[1]{{\bf #1}}

\def\eqa{\begin{eqnarray}}
\def\eea{\end{eqnarray}}
\newcommand{\eq}{\begin{equation}}
\newcommand{\ee}{\end{equation}}

\renewcommand{\>}{\rangle}

\newcommand{\Del}{\Delta}

\usepackage{color}
\begin{document}
	
\title{Comment on ``Unconventional Fermi Surface Instabilities in the Kagome Hubbard Model" by M. Kiesel, C. Platt, and R. Thomale, Phys. Rev. Lett. {\bf 110}, 126405 (2013)}
	
\author{Li-Han Chen}
\affiliation{National Laboratory of Solid State Microstructures \& School of Physics,
	Nanjing University, Nanjing, 210093, China}

\author{Zhen Liu}
\affiliation{National Laboratory of Solid State Microstructures \& School of Physics,
	Nanjing University, Nanjing, 210093, China}

\author{Jian-Ting Zheng}
\email{mf1722040@smail.nju.edu.cn}
\affiliation{National Laboratory of Solid State Microstructures \& School of Physics,
	Nanjing University, Nanjing, 210093, China}
\affiliation{Collaborative Innovation Center of Advanced Microstructures, Nanjing University, Nanjing 210093, China}

\date{\today}% It is always \today, today,
	
\begin{abstract}
This is a comment on Phys. Rev. Lett. {\bf 110}, 126405 (2013), showing it biases the ferromagnetic order more than mean field theories would do. With over-biases like this, the theoretical method applied in the given context is called into question. 
\end{abstract}
	
	
    %\pacs{74.20.-z}{Theories and models of superconducting state}
	%\pacs{74.20.Pq}{Electronic structure calculations}
	%\pacs{74.20.Rp}{Pairing symmetries (other than s-wave)}
	%\pacs{: 74.20.-z, 74.25.Jb, 74.70.Dd}
	%74.70.Xa Pnictides and chalcogenides
	%75.30.Fv  Spin-density waves
	%74.70.Wz  Carbon-based superconductors
	%81.05.ue  Graphene
	%73.22.Pr  Electronic structure of graphene
	%74.20.Rp  Pairing symmetries (other than s-wave)
	%74.20.-z  Theories and models of superconducting state
	%71.27.+a  Strongly correlated electron systems; heavy Fermions
	%64.60.ae  Renormalization-group theory
	%71.10.-w Theories and models of many-electron systems
	%74.62.Dh Effects of cystal defects, doping and substitution (Superconductivity)
	%74.20.Mn : Nonconventional mechanisms
	%74.25.Dw : Superconductivity phase diagrams
	%74.70.-b :  Superconducting materials other than cuprates
	%(for cuprates, see 74.72.-h; for superconducting films, see 74.78.-w)
	
\maketitle

There are several studies on the electronic instabilities in the kagome-Hubbard model at the van-Hove singularity,\cite{thomale,jianxin,qhwang} where the electron density is $5/6$ per site. Surprisingly, Ref.\cite{thomale} reports ferromagnetic (FM) order up to the Hubbard interaction $U=10t$ (when interactions on bonds are absent). Henceforth the nearest-neighbor (NN) hopping $t=1$ is set as the unit of energy. In Refs.\cite{jianxin,qhwang}, other orders are reported. One is the intra-unitcell $120^\circ$ antiferromagnetic (AFM) order. The other is the valence bond solid (VBS) state in the David-star pattern. These orders are illustrated in the insets of Fig.1(c). Here we provide independent checks of all of the above states, showing that FM can not survive at large $U$, while $120^\circ$-AFM and subsequently VBS are more favorable instead.\cite{sm}

In Fig.1(a) we show the Stoner instability lines determined by $U\chi = 1$, where $\chi$ is the bare susceptibility for the respective spin order.\cite{sm} With decreasing temperature $T$, the first instability decides the ordered phase. We see FM only occurs at $U\leq U_0\sim 1.57$ and very low $T$. For all $U>U_0$, the $120^\circ$-AFM order takes over. The phase boundary determined in this way is equivalent to that from the Hatree-Fock mean field theory (HFMFT). To check the reliability, we performed dyanmical mean field theory (DMFT)\cite{dmft} at $T=0$ for the two phases (independently).\cite{sm} The average local spin moment $M$ versus $U$ is shown in Fig.1(b). Consistent with HFMFT, $M$ for FM vanishes for $U> 1$, while $M$ for $120^\circ$-AFM is finite. However, $M$ is reduced by local quantum fluctuations beyond HFMFT.

\begin{figure}
	\includegraphics[width=0.8\columnwidth]{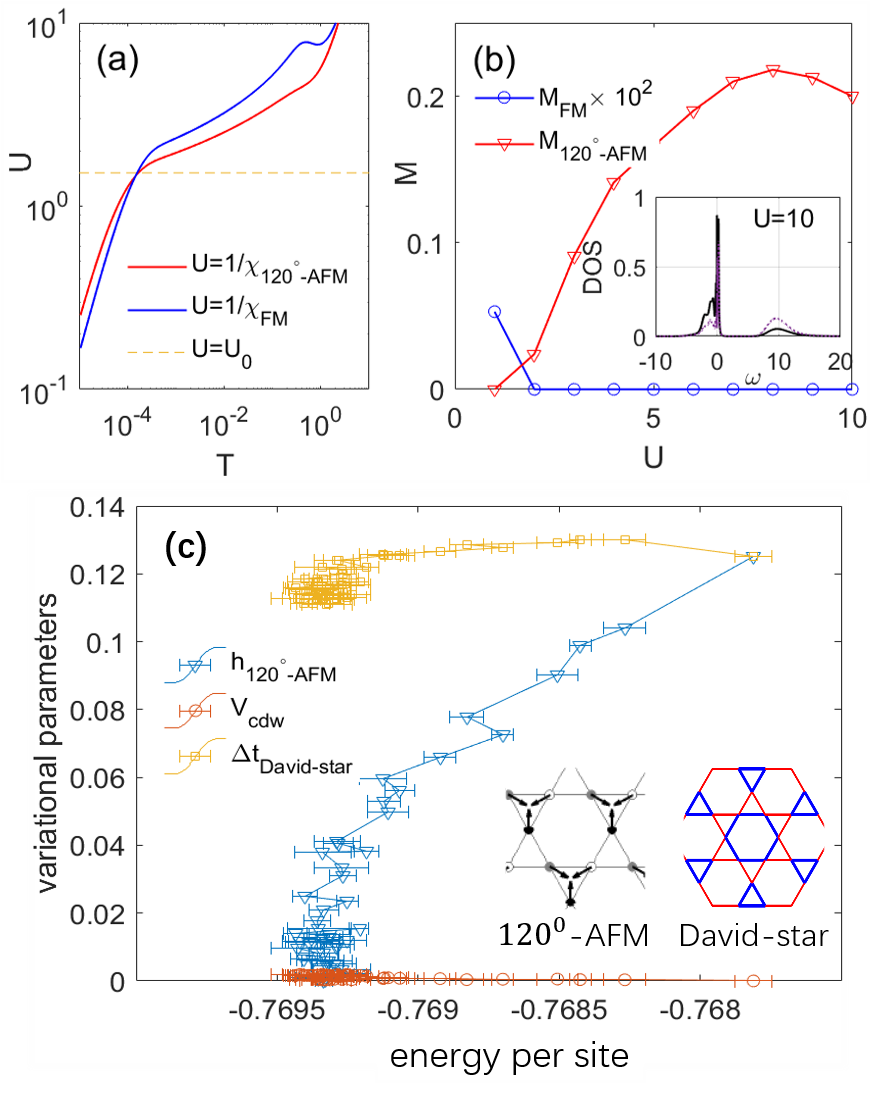}
	\caption{(a) The Stoner instability lines for FM (red) and $120^\circ$-AFM (blue) order. $U_0$ is the crossing point. (b) The average moment from DMFT at zero temperature. The inset shows the spin-diagonal DOS. (c) Trajectory in the energy landscape during the optimization of the variational parameters for an anti-periodic lattice with $6\times 6\times 3$ sites and 90 electrons in total. The left inset shows the $120^\circ$-AFM, and the right one is the David-star VBS, with stronger (weaker) hopping on blue (red) bonds.}\label{fig:mft_vmc}
\end{figure}

At large $U$, say $U=10$, a Mott gap appears in the local density of states (DOS) obtained by DMFT, see the inset of Fig.1(b). At this stage we go beyond DMFT by mapping the Hubbard model to the $t$-$J$ model.\cite{note} We perform variational quantum Monte Carlo in this model.\cite{sm} FM is checked to be unfavorable. This should have been obvious since the NN spin exchange is AFM, and the system is far from the Nagaoka limit. So in the trial many-body wavefunction we include $h_{\rm 120^\circ-AFM}$ to induce the $120^\circ$-AFM, $\Del t_{\rm David-star}$ to enhance (reduce) the hopping on blue (red) bonds, and $V_{\rm cdw}$ to tune the charge-density imbalance on blue hexagons and triangles in the David-star pattern.\cite{sm} The main panel of Fig.1(c) shows the trajectory in the energy landscape as the variational parameters are simultaneously optimized.\cite{vmc} Clearly, the David-star VBS is favorable, while the other parameters vanish as the energy is optimized to the minimum. This also confirms similar results near half filling in Ref.\cite{sigi11,sigi13,sigi14} where spin orders are ignored. 

The fact that FM is absent at large $U$ in MFT acts strongly against its persistence in Ref.\onlinecite{thomale}, since MFT should have over-emphasized any orders within its scope. Therefore, Ref.\onlinecite{thomale} may have biased FM more than MFT would do, to the extent that the $120^\circ$-AFM and David-star VBS at larger $U$ are unfortunately overlooked. With over-biases like this, the theoretical method applied in the given context is called into question.

\acknowledgments{The project was supported by the National Key Research and Development Program of China (under Grant No. 2016YFA0300401), and the National Natural Science Foundation of China (under Grant No.11574134).\\


\begin{references}
\bibitem{note} In the standard form, the $J$-term on bond $\<ij\>$ is $J(\v S_i\cdot \v S_j-n_i n_j/4)$, where $J=4t^2/U$, and $\v S_i$ ($n_i$) is the spin (charge).
\end{references}
\end{document}